\documentclass[12pt,a4]{article}

\usepackage{color,tikz}
\usepackage[unicode,bookmarks,bookmarksopen,bookmarksopenlevel=2,colorlinks,linkcolor=blue,citecolor=green]{hyperref}

\usepackage{amsmath,eucal,amssymb}
\usepackage{mathrsfs,graphicx,texdraw}

\numberwithin{equation}{section}

\def\openone{\leavevmode\hbox{\small1\kern-3.3pt\normalsize1}}

\begin{document}

\begin{center}
{\Large \bf Integrable Discretisations for a Class of Nonlinear Schr\"odinger Equations on Grassmann Algebras}

\bigskip

{\bf Georgi G. Grahovski$^{1,2}$ and  Alexander V. Mikhailov$^1$}

\medskip

{\it $^1$ Department of Applied Mathematics, University of Leeds, Woodhouse Lane,\\ Leeds, LS2 9JT, United Kingdom}\\
{\it $^2$Institute of Nuclear Research and Nuclear Energy,  Bulgarian Academy of Sciences,\\ 72 Tsarigradsko chausee, Sofia 1784, Bulgaria }

\medskip

{\small E-mails: $\quad$  G.Grahovski@leeds.ac.uk, $\quad$ A.V.Mikhailov@leeds.ac.uk}

\end{center}

\begin{abstract}
\noindent
Integrable discretisations for a class of coupled (super) nonlinear Schr\"odinger (NLS) type of equations are presented. The class corresponds to a Lax operator with entries in a Grassmann algebra. Elementary Darboux transformations  are constructed. As a result, Grassmann generalisations of the Toda lattice and the NLS dressing chain are obtained. The compatibility (Bianchi commutativity) of these Darboux transformations leads to integrable Grassmann generalisations of the difference Toda and NLS equations. The resulting  systems will have discrete Lax representations provided by the set of two consistent elementary Darboux transformations.
For the two discrete systems obtained, initial value and initial-boundary problems are formulated.
\end{abstract}

\section{Introduction}\label{sec:1}

There has been a constant interest to noncommutative extensions of integrable equations over the last few decades \cite{Chaish,Ziemek1,DimMH,DimMH3,GrisPen,HaTo,LLPPT,OlvSok1,OlvSok2,Pick,DimMH2,AVM}. Among the well-known examples are the noncommutative analogues of KdV, NLS, sine-Gordon, the KP equation, the Hirota-Miwa equation, two-dimensional Toda lattice equation and AKNS hierarchy.

Supersymmetric  systems are particular examples of noncommutative  extensions of integrable systems. They have attracted much attention because of their applications in physics.  Perhaps the best known example of such equation is the Manin-Radul super-KdV equation \cite{Manin}. It has been extensively studied by various authors \cite{Mathieu,Oevel,AntFor,LiuMan1,LiuMan2,Kupersh}. In \cite{LiuMan1} soliton solutions for the Manin-Radul super-KdV equation were constructed by iterating Darboux transformations, considering the cases of an even number and an odd number of iterations separately. In \cite{LiNim}  unified formulae for the solutions not depending on the parity of the number of iterations are obtained. This is based on an alternative approach to Darboux transformations using quasi-determinants \cite{Gelf2}.

Moreover, the connections between the supersymmetric  soliton equations and the theory of
Lie superalgebras were analysed in \cite{Gurs,MorPi,IK}, in order to obtain a super-analogue
of the classical Drinfeld-Sokolov theory \cite{DrSok*85}.

Darboux transformations play a crucial role in the theory of integrable systems \cite{Matv}. They are gauge-like transformations which preserve the form of the associated linear problem (Lax representation) and  are basic tools for obtaining the so-called soliton solutions \cite{brown-bible,zm1,DimMH,Ciesh2}. The theory of Darboux transformations was boosted by the dressing method \cite{shabat}.

Darboux transformations also play a role in constructing integrable discretisations of integrable equations \cite{Adl}: to a very wide variety of (continuous) integrable nonlinear PDEs the associated dressing chains can be interpreted as B\"acklund-Darboux transformations \cite{Levi}. Furthermore, the compatibility (Bianchi commutativity) of these B\"acklund-Darboux transformations leads to equations with two discrete independent variables (lattice equations). The Bianchi commutativity for B\"acklund-Darboux transformations is also known as a principle for nonlinear superposition \cite{Levi}. The classifications of elementary Darboux transforms can be used as a tool to classify discrete systems related to a given Lax operator. These discrete systems will have Lax pairs provided by the set of two consistent Darboux transformations. The corresponding B\"acklund transformations will represent symmetries of the discrete (difference systems).

Here we will study Darboux transformation for a class of the nonlinear Schr\"odinger (NLS) type of systems with Lax pairs belonging to  Grassmann algebras. Such systems are closely-related to the supersymmetric version of the NLS equation with an $osp(1|2)$-invariant Lax pair studied by P. P. Kulish \cite{Kulish1}:
\begin{equation}\label{eq:1}
\begin{split}
{\rm i} u_t + u_{xx} - 2u^\dag u u - \Psi^\dag \Psi u + {\rm i} \Psi \Psi_x=0\\
{\rm i} \Psi_t + \Psi_{xx} - u^\dag u \Psi + {\rm i} (2u\Psi^\dag_x + \Psi^\dag u_x)=0
\end{split},
\end{equation}
Here  $u$ and $\Psi$ are smooth functions taking values in a  Grassmann algebra ${\mathcal G}= {\mathcal G}_0 \oplus {\mathcal G}_1$. The variables $u$ are called  commuting (bosonic) variables: $u_1 u_2 = u_2 u_1$, $u_1, u_2 \in {\mathcal G}_0$ while the variables $\Psi$ are called anti-commuting (fermionic) ones: $\Psi_1 \Psi_2 = -\Psi_2 \Psi_1$, $\Psi_1, \Psi_2 \in {\mathcal G}_1$. Here also $\dag$ stays for  a Hermitian conjugation in ${\mathcal G}$.

The paper is organised as follows: in Section 2  we present some basic facts of Grassmann algebras, the Lax representation and the general form of the partial differential equations. In Section 3 we describe the elementary Darboux transformations and the corresponding dressing chains. In Section 4 we present the discrete integrable systems obtained as a compatibility condition of elementary Darboux transformations from Section 3.

\section{Preliminaries: Grassmann algebras and Lax representation}\label{sec:2}

Firstly, we present  basic facts and properties of Grassmann algebras. Further details can be found e.g. in \cite{Ber,Leites}.

Let ${\mathcal G}$ be a ${\Bbb Z}_2$-graded algebra over a field $K$ of characteristics zero (such as ${\Bbb C}$ or ${\Bbb Q}$). Thus,  ${\mathcal G}$ as a linear space is a direct sum ${\mathcal G}={\mathcal G}_0 \oplus {\mathcal G}_1$,  such that ${\mathcal G}_i{\mathcal G}_j \subseteq {\mathcal G}_{i+j} \quad ({\rm mod} \, 2)$. Those elements of ${\mathcal G}$ that belong either to ${\mathcal G}_0$ or to ${\mathcal G}_1$ are called homogeneous, the ones from ${\mathcal G}_0$ are called even, while those in ${\mathcal G}_1$ are called odd.

By definition, the parity $|a|$ of an even homogeneous element $a$ is $0$ and it is $1$ for odd homogeneous elements. The parity of the product $|ab|$ of two homogeneous elements is a sum of their parities: $|ab|= |a|+|b|$.
Grassmann commutativity means that $ba=(-1)^{|a||b|}ab$ for any homogenous elements $a$ and $b$. In particular, $a_1^2=0$, for all $a_1\in {\mathcal G}_1$ and even elements commute with all elements of ${\mathcal G}$.

Consider a Lax operator of the form
\begin{eqnarray}\label{eq:Lax}
L= \partial_x+U -\lambda h,
\end{eqnarray}
where the matrix $U$ has entries in a Grassmann algebra:
\begin{eqnarray}\label{eq:U}
U= \left( \begin{array}{ccc} 0 & \psi & 2q \\ -\varkappa & 0 & \zeta \\ 2p & \phi & 0 \\\end{array}\right), \qquad h={1\over 2}\left( \begin{array}{ccc} 1 & 0 & 0 \\ 0 & 0 & 0 \\ 0 & 0 & -1 \\\end{array}\right).
\end{eqnarray}
Here
 $p$ and $q$ are even elements of the Grassmann algebra ${\mathcal G}$, while $\zeta$, $\varkappa$, $\phi$ and $\psi$ are odd homogeneous ones and $\lambda \in {\Bbb C}$ is a spectral parameter (even variable). We will be using the natural grading $U_{ij}\in {\mathcal G}_{i+j}$ (mod $2$).

The zero curvature condition  $[L,A]=0$ leads to the following system of  equations:
\begin{eqnarray}\label{eq:nlee}
q_t&=&-q_{xx} + \psi_{x}\zeta - \psi\zeta_{x} - 2q(\psi\varkappa - \phi \zeta) + 8 q^2p\nonumber\\
p_t&=&p_{xx} + \phi_{x}\varkappa - \phi\varkappa_{x} + 2p(\psi\varkappa - \phi \zeta) - 8 p^2q\nonumber\\
\psi_{t}&=&\psi_{xx} - q_x\phi - 2q\phi_{x} + 2 pq \psi + \psi\phi\zeta\\
\zeta_{t}&=&\zeta_{xx} - q_x\varkappa - 2q\varkappa_{x} + 2 pq \zeta - \psi\varkappa\zeta\nonumber\\
\varkappa_{t}&=&-\varkappa_{xx} + p_x\zeta + 2p\zeta_{x} + 2 pq \varkappa - \phi\varkappa\zeta\nonumber\\
\phi_{t}&=&-\phi_{xx} + p_x\psi + 2p\psi_{x} + 2 pq \phi + \psi\phi\varkappa\nonumber
\end{eqnarray}
This is a super-analogue of the NLS equation. It differs from the supersymmetric AKNS problem \cite{Gurs}. The second Lax operator $A$ is of the form:
\begin{eqnarray}\label{eq:Lax-M}
A= \partial_t+V_0 + \lambda U -\lambda^2 h,
\end{eqnarray}
where
\begin{eqnarray}\label{eq:V}
V_0= \mbox{ad}_h^{-1} \, U_x + \left( \begin{array}{ccc} 4pq-2\psi \varkappa & 0 & 0 \\ 0 & -4pq - 2 \phi \zeta & 0 \\ 0 & 0 & 2(\phi \zeta + \psi \varkappa) \\\end{array}\right).
\end{eqnarray}
The reduction $\psi=\zeta=\phi^\dag=\varkappa^\dag$ and $p=q^\dag$ leads to a system which after a re-scaling and a point transformation $t\rightarrow {\rm i}t$, $x\rightarrow {\rm i}x$  leads to (\ref{eq:1}). It can be shown that the system (\ref{eq:nlee}) is a completely integrable Hamiltonian system. It has an infinite number of conservation laws. The first three constants of motion are of the form:
\begin{eqnarray}\label{eq:nlee-ints}
\mathcal{N} &= \int_{-\infty}^\infty {\rm d}x\, &\{4pq + \phi\psi+ \varkappa \zeta \}; \\
\mathcal{P} &= \int_{-\infty}^\infty {\rm d}x\, &\{-2pq - \phi\psi_x- \varkappa \zeta_x- \phi_x\psi- \varkappa_x\zeta\};\\
\mathcal{H} &= \int_{-\infty}^\infty {\rm d}x\, &\{2p_x q_x + \phi_x\psi_x+ \varkappa_x\zeta_x + 2p^2q^2 + pq(\phi \psi + \varkappa \zeta) \nonumber\\
&&   -q (\phi \phi_x +\varkappa \varkappa_x )-p(\psi \psi_x + \zeta\zeta_x)\}.
\end{eqnarray}
In physical applications $\mathcal{N}$ is known as a "total number of particles", $ \mathcal{P} $ as a "total momentum" while $ \mathcal{H} $ is the Hamiltonian of the system.

\section{Darboux transforms for the Lax operator (\ref{eq:Lax})}\label{sec:3}

By a Darboux transformation we understand a map
\begin{eqnarray}\label{eq:DT}
L \rightarrow L_1=MLM^{-1}
\end{eqnarray}
where the Lax operator $L_1$ has the same form (\ref{eq:Lax}) but with an updated potential $U_1$:
\begin{eqnarray}\label{eq:Lax1}
L_1= \partial_x+U_1 -\lambda h, \qquad U_1= \left( \begin{array}{ccc} 0 & \psi_1 & 2q_1 \\ -\varkappa_1 & 0 & \zeta_1 \\ 2p_1 & \phi_1 & 0 \\\end{array}\right).
\end{eqnarray}
In (\ref{eq:DT}) $M$ is  a matrix whose entries $M_{ij}\in {\mathcal G}_{i+j}$ (mod 2) are rational functions of $\lambda$ and differentiable functions of $x$. It follows from (\ref{eq:DT}) that
\begin{eqnarray}\label{eq:Mcomp}
M_x + U_1M- MU - \lambda [h,M]=0.
\end{eqnarray}
Obviously, a composition of Darboux transformations is again a Darboux transformation with more complicated rational dependence in $\lambda$.

We are interested in elementary Darboux transformations which cannot be decomposed further. Thus, we restrict ourselves by linear in $\lambda$ Darboux matrices:
\begin{eqnarray}\label{eq:M}
M= M_0 + \lambda M_1.
\end{eqnarray}
The substitution of $M$ in (\ref{eq:Mcomp}) results in:
\begin{eqnarray}\label{eq:MS-a}
[h,M_1]=0,\\
M_{1,x} + U_1M_1- M_1U - [h,M_0]=0,\label{eq:MS-b}\\
M_{0,x} + U_1M_0- M_0U=0.\label{eq:MS-c}
\end{eqnarray}
Let us consider the simplest case of $\lambda$- independent Darboux transformations ($M_1=0$).   From (\ref{eq:MS-b}) it follows that $M_0$ is a diagonal matrix. Then, equation (\ref{eq:MS-c}) implies that $M_0$ is a constant diagonal matrix, and that $U_1M_0=M_0U$. The later is nothing but a  Lie point symmetry transformation, which does not lead to nontrivial results.

If $M_1\neq 0$, then it follows from (\ref{eq:MS-a}) that $M_1$ is a diagonal matrix: $M_1=\mbox{diag} \, (\alpha, \beta,\gamma)$. Furthermore, the equation (\ref{eq:MS-b}) implies that $\alpha$, $\beta$ and $\gamma$ are constants. We will describe the elementary Darboux transformations. In this case the matrix $M_1$ in (\ref{eq:M}) has rank one and without any loss of generality, we can set $\alpha =1$, $\beta = 0$, $\gamma = 0$. It follows from (\ref{eq:MS-b}) and (\ref{eq:MS-c}) that $M_{0,22}$ and $M_{0,33}$ are constants. Further analysis shows that there are two essentially different cases: (1) $M_{0,22}=1$ and $M_{0,33}=0$; (2) $M_{0,22}=M_{0,33}=1$.

For a sake of convenience, from now on, we will denote the matrix element $M_{11}$ by $F$.

{\bf Case 1:} $M_{22}=1$ and $M_{33}=0$.
The $\lambda$-term of the compatibility condition (\ref{eq:MS-b}) in this case gives the conditions: $M_{12}=\psi$, $M_{21}=-\varkappa_1$, $M_{13}=p$, $M_{31}=p_1$ and $M_{23}=M_{32}=0$.
Therefore, the Darboux matrix $M$ takes the form
\begin{eqnarray}\label{eq:DT1}
M=\left( \begin{array}{ccccc} F +\lambda& \, & \psi &\, & q \\ - \varkappa_1 &\, & 1 &\, & 0 \\ p_1 &\, & 0 &\, & 0 \\\end{array}\right).
\end{eqnarray}
Then, the $\lambda$-independent term in the compatibility condition (\ref{eq:MS-c}) leads to
the set of algebraic constrains:
\begin{eqnarray}\label{eq:DT1constr}
\phi_1=-p_1\psi, \qquad \zeta=-q\varkappa_1, \qquad p_1q=1;
\end{eqnarray}
and to the following system of equations:
\begin{eqnarray}\label{eq:DT1chain}
q_x &=&- (\psi \varkappa_1+2F )  q \nonumber\\
F_{x}&=&2\left({q_1\over q} -{q\over q_{-1}}\right) + \psi\varkappa - \psi_1\varkappa_1  \nonumber\\
\psi_{x}&=&\psi_1 -\psi F + {q\over q_{-1}}\psi_{-1} \\
\varkappa_{1,x}&=&-\varkappa +\varkappa_1F + {q_1\over q}\varkappa_2\nonumber
\end{eqnarray}
Now, if we introduce new variables $v$, $\phi$ and $\psi$ together with the  forward $v_1$ and backward $v_{-1}$ shifts of $v$
\[
q={\rm e}^{v}, \qquad p= {\rm e}^{v_{-1}}, \qquad \psi=\eta {\rm e}^{v/2}, \qquad \varkappa_1=\varphi {\rm e}^{-v/2},
\]
using also the algebraic constrains (\ref{eq:DT1constr}) one can eliminate the function $F$ and  cast (\ref{eq:DT1chain}) into a system of 3 equations:
\begin{eqnarray}\label{eq:DT1Toda}
v_{xx}&=&4\left({\rm e}^{v_1-v} - {\rm e}^{v-v_{-1}}\right) + (\varphi \eta_{-1} + \varphi_{-1}\eta ){\rm e}^{(v-v_{-1})/2} - (\varphi \eta_{1} + \varphi_{1}\eta ){\rm e}^{(v_1-v)/2}\nonumber\\
\varphi_{x}&=&\varphi_{1} {\rm e}^{(v_1-v)/2} + \varphi_{-1} {\rm e}^{(v-v_{-1})/2}\\
\eta_{x}&=&-\eta_{1} {\rm e}^{(v_1-v)/2} - \eta_{-1} {\rm e}^{(v-v_{-1})/2}\nonumber
\end{eqnarray}
The above system is an integrable noncommutative extension of the Toda chain: the reduction $\xi=\eta=0$ leads to the standard Toda chain:
\[
v_{xx}=4{\rm e}^{v_1-v} - 4 {\rm e}^{v-v_{-1}}.
\]
The system (\ref{eq:DT1Toda}) also has a Lagrangian formulation with a Lagrangian:
\begin{eqnarray}\label{eq:DT1lagr}
\mathcal{L}(v, \xi, \eta) = \int {\rm d}x \, \left({v_x^2\over 2} - 4 {\rm e}^{v-v_{-1}} + 2(\varphi \eta_{-1} + \varphi_{-1}\eta ){\rm e}^{(v-v_{-1})/2}+ \varphi \eta_x - \varphi_x \eta \right).
\end{eqnarray}

{\bf Case 2:} $M_{22}=1$, $M_{33}=1$.

It follows from(\ref{eq:MS-b}) that  $M_{12}=\psi$, $M_{21}=-\varkappa_1$, $M_{13}=q$, $M_{31}=p_1$ and $M_{23}=M_{32}=0$.
Due to Abel's theorem, the  Wronskian  does not depend on $x$ (since the potential $U$ is a traceless matrix) and thus $(F-  p_1q + \psi\varkappa_1)_x=0$. Therefore, the function $F$ can be determined up to a constant $\mu \in {\mathcal G}_0$:
\begin{eqnarray}\label{eq:FI1}
F=  p_1q - \psi\varkappa_1 + \mu.
\end{eqnarray}
 As a result, the Darboux matrix $M$ takes the form
\begin{eqnarray}\label{eq:DT2}
M=\left( \begin{array}{ccccc} \mu + p_1q - \psi\varkappa_1 +\lambda& \, & \psi &\, & q \\ - \varkappa_1 &\, & 1 &\, & 0 \\ p_1 &\, & 0 &\, & 1 \\\end{array}\right).
\end{eqnarray}
Then, the condition (\ref{eq:MS-c}) leads to the following  algebraic equations:
\begin{eqnarray}\label{eq:DT2constr}
(1-T)\zeta=(q \varkappa_1), \qquad (1-T)\phi=(p_1 \psi).
\end{eqnarray}
and to a set of 4 dressing chain equations:
\begin{eqnarray}\label{eq:DT2chain}
q_x &=& 2q (\psi\varkappa_1 - p_1q - \mu) + 2 q_1 - (1-T)^{-1}(q\varkappa_1)\psi\nonumber\\
p_{x}&=& -2p (\psi_{-1}\varkappa - pq_{-1} - \mu) - 2 p_{-1} - (1-T)^{-1}(p_1\psi)\varkappa\nonumber\\
\psi_{x}&=&\psi_1 - q(1-T)^{-1}(p_1 \psi) -(\mu + p_1q) \psi \\
\varkappa_{x}&=& -\varkappa_{-1} - p(1-T)^{-1}(q\varkappa_1) +(\mu + pq_{-1})\varkappa \nonumber
\end{eqnarray}
Here $T$ is the shift operator induced by the Darboux transform (\ref{eq:DT2}). It acts on the potential as  $U_1=TU =\mbox{Ad}_{M_1} U$: $q_1=Tq$, $p_1=Tp$, $\psi_1=T\psi$, $\phi_1=T\phi$, $\zeta_1=T\zeta$ and $\varkappa_1=T\varkappa$. The presence of the operator $(1-T)^{-1}$ in (\ref{eq:DT2chain}) leads to a nonlocal dressing chain. We should note that (\ref{eq:DT2chain}) can be rewritten into a local form in  terms of the variables $p$, $q$, $\phi$ and $\zeta$, although  it will lead to non-evolutionary dressing chain equations for the odd variables $\phi$ and $\zeta$.

In the bosonic limit (when all noncommuting variables vanish), (\ref{eq:DT2chain}) reduces to the standard NLS dressing chain \cite{Adl}:
\begin{eqnarray}\label{eq:NLSdress}
q_x = -2q (p_1q + \mu) + 2 q_1 \qquad
p_{1,x}= 2p_1 (p_1q + \mu) - 2 p.
\end{eqnarray}

\section{Darboux transforms and discretisation}\label{sec:4}

A standard approach for obtaining integrable difference equations is to construct a pair of Darboux transformations: discrete systems appear as consistency conditions of two Darboux matrices $M$ and $N$ around a square \cite{Levi,Adl} (the Bianchi commutativity).

Let us introduce lattice variables $(k,m)$ ($k,m\in {\Bbb Z}$): generic even  $v_{k,m}$ and odd  variables $\tau_{k,m}$ are defined on an integer lattice ${\Bbb Z}\times {\Bbb Z}$ (through the whole text we adopt the convention to denote by Greek letters the odd variables and with Latin letters the even ones). It is also useful to introduce the shift operators $S$ and $T$ as follows: the operator $S$ shifts the first lattice variable while the operator $T$ shifts the second one. For example
\begin{eqnarray}\label{eq:Shifts}
Sq_{k,m}=q_{k+1,m}, \qquad T\zeta_{k,m}=\zeta_{k,m+1}, \qquad TS=ST.
\end{eqnarray}
Now, consider two Darboux transformations $M(\lambda)$ and $N(\lambda)$. On the space of fundamental solutions $\{\Psi\}$ of $L(\lambda)$ (\ref{eq:Lax}) they act as follows:
\begin{eqnarray}\label{eq:FS}
S[\Psi(\lambda)]= M(\lambda)\Psi(\lambda),\qquad
T[\Psi(\lambda)]= N(\lambda)\Psi(\lambda).
\end{eqnarray}
The compatibility of the transformations $[S,T]=0$ (Bianchi commutativity) implies
\begin{eqnarray}\label{eq:DTconsist}
S[N(\lambda)]M(\lambda)=T[M(\lambda)]N(\lambda),
\end{eqnarray}
and leads to a set of algebraic relations between $U$, $S[U]$, $T[U]$ and $TS[U]$.

In this setting,  Darboux transformations (\ref{eq:FS}) can be considered as a discrete Lax pair  associated with $L(\lambda)$, while the system of algebraic equations (\ref{eq:DTconsist}) can be considered as a system of difference equations for variables labelled by the point on the lattice. Hence, the system (\ref{eq:DTconsist}) is an integrable discretisation of the hierarchy of $L(\lambda)$ having a discrete Lax pair given by (\ref{eq:FS}). Moreover, the differential equations for the B\"acklund transformation, coming from the derivation of the Darboux matrices can be considered as symmetries of the difference system.

Here we will describe the set of integrable discretisations obtained from imposing a consistency of two Darboux transformations, obtained in the previous Section.

{\bf Case A:} In this case, consider two Darboux matrices of the type described in Case 2 in section \ref{sec:3}:
\begin{eqnarray}\label{eq:DDT2}
M=\left( \begin{array}{ccccc} \mu + p_{10}q - \psi\varkappa_{10} +\lambda& \, & \psi &\, & q \\ - \varkappa_{10} &\, & 1 &\, & 0 \\ p_{10} &\, & 0 &\, & 1 \\\end{array}\right); \quad
N=\left( \begin{array}{ccccc} \nu + p_{01}q - \psi\varkappa_{01} +\lambda& \, & \psi &\, & q \\ - \varkappa_{01} &\, & 1 &\, & 0 \\ p_{01} &\, & 0 &\, & 1 \\\end{array}\right),
\end{eqnarray}
The consistency condition (\ref{eq:DTconsist}) leads to the following discrete system of four equations:
\begin{eqnarray}\label{eq:Discr1}
p_{01} - p_{10} &=&{\mu-\nu \over (1 + p_{11}q)^2}(1 + p_{11}q + \psi\varkappa_{11})p_{11} \nonumber\\
q_{01} - q_{10} &=&-{\mu-\nu \over (1 + p_{11}q)^2}(1 + p_{11}q + \psi\varkappa_{11})q \nonumber\\
\varkappa_{01} - \varkappa_{10} &=&{\mu-\nu \over 1 + p_{11}q}\varkappa_{11}\\
\psi_{01} - \psi_{10} &=&-{\mu-\nu \over 1 + p_{11}q}\psi \nonumber
\end{eqnarray}
If all odd variables vanish, this system of difference equations reduced to a familiar two-component system \cite{Adler1,Adler2,Adl}:
\begin{eqnarray}\label{eq:Discr1r}
p_{01} - p_{10} ={(\mu-\nu) p_{11}\over 1 + p_{11}q} \qquad
q_{01} - q_{10} =-{(\mu-\nu)q \over 1 + p_{11}q}.
\end{eqnarray}

\begin{figure}[htb]
\begin{center}
\begin{tikzpicture}
\filldraw [black] (0,0) circle (2pt)
(0,1) circle (2pt)
(1,1) circle (2pt)
(1,2) circle (2pt)
(2,2) circle (2pt)
(2,3) circle (2pt)
(3,3) circle (2pt)
(3,4) circle (2pt)
(4,4) circle (2pt)
(4,5) circle (2pt);

\draw [-, thick](0,0) -- (-0.5,0);
\draw [-, thick](0,0) -- (0,1);
\draw [-, thick](0,1) -- (1,1);
\draw [-, thick](1,1) -- (1,2);
\draw [-, thick](1,2) -- (2,2);
\draw [-, thick](2,2) -- (2,3);
\draw [-, thick](2,3) -- (3,3);
\draw [-, thick](3,3) -- (3,4);
\draw [-, thick](3,4) -- (4,4);
\draw [-, thick](4,4) -- (4,5);
\draw [-, thick](4,5) -- (4.5,5);
\draw [dashed, thick](0,0) -- (1,0);
\draw [dashed, thick](1,0) -- (1,1);
\draw [dashed, thick](1,1) -- (2,1);
\draw [dashed, thick](2,1) -- (2,2);
\draw [dashed, thick](2,2) -- (3,2);
\draw [dashed, thick](3,2) -- (3,3);
\draw [dashed, thick](3,3) -- (4,3);
\draw [dashed, thick](4,3) -- (4,4);
\draw [dashed, thick](4,4) -- (4.5,4);
\draw [dashed, thick](0,0) -- (0,-0.5);
\draw [dashed, thick](0,1) -- (-0.5,1);
\draw [dashed, thick](0,1) -- (0,2);
\draw [dashed, thick](0,2) -- (1,2);
\draw [dashed, thick](1,2) -- (1,3);
\draw [dashed, thick](1,3) -- (2,3);
\draw [dashed, thick](2,3) -- (2,4);
\draw [dashed, thick](2,4) -- (3,4);
\draw [dashed, thick](3,4) -- (3,5);
\draw [dashed, thick](3,5) -- (4,5);
\draw [dashed, thick](4,5) -- (4,5.5);

\draw [->, thick](-2.5,0) -- (-2.5,1);
\draw [->, thick](-2.5,0) -- (-1.5,0);

\draw [->, thick](1,5) -- (-2,5);
\draw [->, thick](3,0.5) -- (6,0.5);

\draw (-0.5,4.5) node {($p_{01}$, $q_{01}$, $\psi_{01}$, $\varkappa_{01}$)};
\draw (4.5,1) node {($p_{10}$, $q_{10}$, $\psi_{10}$, $\varkappa_{10}$)};
\draw (-2.5,1.25) node {$m$};
\draw (-1.25,0) node {$k$};
\end{tikzpicture}
\end{center}
\caption{{\em{A staircase initial value problem for (\ref{eq:Discr1})}}} \label{fig:2}
\end{figure}

One can easily solve (\ref{eq:Discr1}) either with respect to ($p_{10}$, $q_{10}$, $\psi_{10}$, $\varkappa_{10}$) or with respect to  ($p_{01}$, $q_{01}$, $\psi_{01}$, $\varkappa_{01}$). In this case, one can pose an initial value problem for  (\ref{eq:Discr1}) with initial conditions on a staircase as it is shown on Figure \ref{fig:2}.   For a given set of initial data on the staircase, a solution of the difference system (\ref{eq:Discr1}) can be found recursively.

Similar to \cite{MWX2} one can define the Elimination map and express any variable on the lattice in terms of a finite subset of the initial set of variables on the staircase  (Fig. \ref{fig:2}). It is clear, that these expressions are rational functions of the even initial variables and multi-linear function of the odd ones.

{\bf Case B:} If we combine two Darboux transformations of types (\ref{eq:DT2}) and (\ref{eq:DT1}) from the previous section:
\begin{eqnarray}\label{eq:DDT22}
M=\left( \begin{array}{ccccc} \mu + p_{10}q - \psi\varkappa_{10} +\lambda& \, & \psi &\, & q \\ - \varkappa_{10} &\, & 1 &\, & 0 \\ p_{10} &\, & 0 &\, & 1 \\\end{array}\right), \quad N=\left( \begin{array}{ccccc} F +\lambda& \, & \psi &\, & q \\ - \varkappa_{01} &\, & 1 &\, & 0 \\ p_{01} &\, & 0 &\, & 0 \\\end{array}\right);
\end{eqnarray}
then the compatibility (the Bianchi commutativity) condition will lead to the following quadrilateral system:
\begin{eqnarray}\label{eq:Discr2-a}
p_{01}&=&(\mu -F + p_{10}q - \psi\varkappa_{10})p_{11} \\
q_{10} &=&(\mu -F_{10} + p_{11}q_{01} - \psi_{01}\varkappa_{11})q \label{eq:Discr2-b}\\
\psi_{01}-\psi_{10}&=&-(\mu -F_{10} + p_{11}q_{01} - \psi_{01}\varkappa_{11})\psi \label{eq:Discr2-c}\\
\varkappa_{01}-\varkappa_{10}&=&(\mu -F + p_{10}q - \psi\varkappa_{10})\varkappa_{11} \label{eq:Discr2-d}\\
F(\mu + p_{11}q_{01}-\psi_{01}\varkappa_{11}) &=& F_{10}(\mu + p_{10}q-\psi\varkappa_{10}) + p_{10}q_{10} - p_{01}q_{01} - \psi_{10}\varkappa_{10} + \psi_{01}\varkappa_{01},\label{eq:Discr2-f}
\end{eqnarray}
and the condition: $p_{01}q=1$ (\ref{eq:DT1constr}) which enables us to eliminate the variable $p$. One can solve (\ref{eq:Discr2-a}) and (\ref{eq:Discr2-b}) with respect to $F$ and its shift $F_{10}$:
\begin{eqnarray}\label{eq:Discr2-F}
F=\mu - {q_{10}\over q} + {q\over q_{1,-1}} - \psi \varkappa_{10}, \qquad F_{10}=\mu - {q_{01}\over q} + { q_{01}\over q_{10}} - \psi_{01} \varkappa_{11}.
\end{eqnarray}
Then, using (\ref{eq:Discr2-F}), one can eliminate $F$ from (\ref{eq:Discr2-c}) and (\ref{eq:Discr2-d}). The compatibility condition $S(F)=F_{10}$ and (\ref{eq:Discr2-c})-(\ref{eq:Discr2-d}) read:
\begin{eqnarray}\label{eq:Discr2rA}
{q\over q_{-1,0}} + {q\over q_{1,-1}} -{q_{-1,1}\over q} - {q_{1,0}\over q} + \psi_{-1,1} \xi - \psi \xi_{1,-1} + \mu -\mu_1 =0, \\
 \psi_{1,0}-\psi_{01}= {q_{10}\over q} \psi,\label{eq:Discr2rA-odd1}\\
\xi-\xi_{1,-1}= {q_{10}\over q}\xi_{10}.\label{eq:Discr2rA-odd2}
\end{eqnarray}
Here we have introduced a new variable  $\xi=\varkappa_{0,1}$. The equation (\ref{eq:Discr2-f}) is satisfied due to (\ref{eq:Discr2-F}) and (\ref{eq:Discr2rA})-(\ref{eq:Discr2rA-odd2}).

After setting $q={\rm e}^{v}$, where $v$ is an even variable, one can easily recognise (\ref{eq:Discr2rA}) as a noncommutative extension of the discrete Toda lattice:
\begin{eqnarray}\label{eq:Discr2r-TCa}
{\rm e}^{v_{1,-1}-v} + {\rm e}^{v_{-1,0}-v}- {\rm e}^{v-v_{1,0}} - {\rm e}^{v-v_{-1,1}} + \psi_{-1,1} \xi - \psi \xi_{1,-1}   + \mu_1 -\mu =0,\\
 \psi_{1,0}-\psi_{01}= {\rm e}^{v_{1,0}-v} \psi,\label{eq:Discr2rA-odd1a}\\
\xi-\xi_{1,-1}= {\rm e}^{v_{1,0}-v}\xi_{10}.\label{eq:Discr2rA-odd2a}
\end{eqnarray}
In the special case when all anti-commuting variables vanish, it reduces to  the  discrete Toda lattice \cite{Chen,Toda}:
\begin{eqnarray}\label{eq:Discr2r-TC}
{\rm e}^{v_{1,-1}-v} + {\rm e}^{v_{-1,0}-v}- {\rm e}^{v-v_{1,0}} - {\rm e}^{v-v_{-1,1}}  + \mu_1 -\mu =0.
\end{eqnarray}
On the lattice each equation (\ref{eq:Discr2rA}) - (\ref{eq:Discr2rA-odd2}) can be represented by a graph (stencil) as follows:
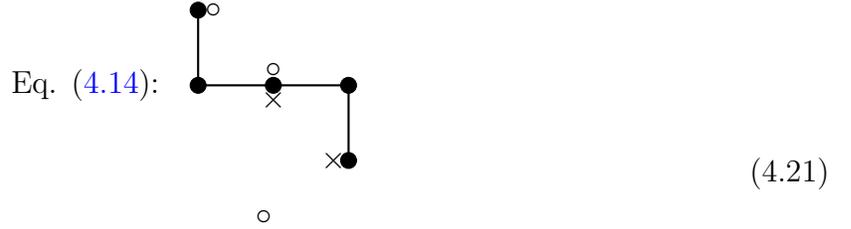
\begin{eqnarray}\label{eq:elem_ev-q}
 \begin{tikzpicture}
\filldraw [black] (0,0) circle (3pt)
(1,0) circle (3pt)
(1,-1) circle (3pt)
(-1,0) circle (3pt)
(-1,1) circle (3pt);
\draw [-,black, thick](0,0) -- (1,0);
\draw [-,black, thick](-1,1) -- (-1,0);
\draw [-,black, thick](-1,0) -- (0,0);
\draw [-,black, thick](1,0) -- (1,-1);
\draw [black](0,0.2) node {$\circ$};
\draw [black](-0.8,1) node {$\circ$};
\draw [black](0,-0.2) node {$\times$};
\draw [black](0.8,-1) node {$\times$};
\draw [black](-2.5,0) node {Eq. (\ref{eq:Discr2rA}):};
\end{tikzpicture}
\\
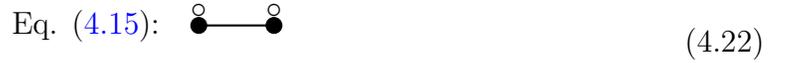
\begin{tikzpicture}
\filldraw [black] (0,0) circle (3pt)
(1,0) circle (3pt);
\draw [-,black, thick](0,0) -- (1,0);
\draw [black](0,0.2) node {$\circ$};
\draw [black](1,0.2) node {$\circ$};
\draw [black](0,1) node {$\circ$};
\draw [black](-1.5,0) node {Eq. (\ref{eq:Discr2rA-odd1}):};
\end{tikzpicture} \\
 
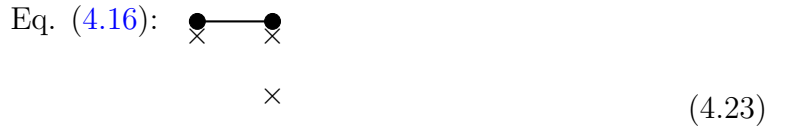
\begin{tikzpicture}
\filldraw [black] (0,0) circle (3pt)
(1,0) circle (3pt);
\draw [-,black, thick](0,0) -- (1,0);
\draw [black](0,-0.2) node {$\times$};
\draw [black](1,-0.2) node {$\times$};
\draw [black](1,-1) node {$\times$};
\draw [black](-1.5,0) node {Eq. (\ref{eq:Discr2rA-odd2}):};
\end{tikzpicture}\label{eq:elem_ev-xi}
\end{eqnarray}
where even variables are denoted by solid dots \begin{tikzpicture}
\filldraw [black] (0,0) circle (3pt);\end{tikzpicture}, the variables $\psi$ -- with empty circles $\circ$ and the variables $\xi$ with crosses $\times$. To each ratio in (\ref{eq:Discr2rA}) there corresponds a solid line.

\begin{figure}[htb]
\begin{center}
\begin{tikzpicture}
\filldraw [black] (4,4) circle (3pt)
(3,4) circle (3pt)
(4,3) circle (3pt)
(5,3) circle (3pt)
(5,2) circle (3pt)
(6,2) circle (3pt)
(6,1) circle (3pt)
(7,1) circle (3pt)
(7,0) circle (3pt)
(3,5) circle (3pt)
(2,5) circle (3pt)
(2,6) circle (3pt)
(1,6) circle (3pt)
(1,7) circle (3pt)
(0,7) circle (3pt)
(8,0) circle (3pt)
(8,-1) circle (3pt)
(0,8) circle (3pt);
\draw [-,black, thick](0,8) -- (-0.5,8);
\draw [-,black, thick](0,8) -- (0,7);
\draw [-,black, thick](0,7) -- (1,7);
\draw [-,black, thick](1,7) -- (1,6);
\draw [-,black, thick](1,6) -- (2,6);
\draw [-,black, thick](2,6) -- (2,5);
\draw [-,black, thick](2,5) -- (3,5);
\draw [-,black, thick](3,5) -- (3,4);
\draw [-,black, thick](3,4) -- (4,4);
\draw [-,black, thick](4,4) -- (4,3);
\draw [-,black, thick](4,3) -- (5,3);
\draw [-,black, thick](5,3) -- (5,2);
\draw [-,black, thick](5,2) -- (6,2);
\draw [-,black, thick](6,2) -- (6,1);
\draw [-,black, thick](6,1) -- (7,1);
\draw [-,black, thick](7,1) -- (7,0);
\draw [-,black, thick](7,0) -- (8,0);
\draw [-,black, thick](8,0) -- (8,-1);
\draw [-,black, thick](8,-1) -- (8.5,-1);
\draw [-,gray, thick](8.1,-1.1) -- (8.1,4.1);
\draw [-,gray, thick](8.1,-1.1) -- (-0.1,2.9);
\draw [-,gray, thick](-0.1,2.9) -- (-0.1,8.1);
\draw [-,gray, thick](-0.1,8.1) -- (8.1,4.1);

\draw [dashed](-0.5,8) -- (0.5,8);
\draw [dashed](-0.5,7) -- (2.5,7);
\draw [dashed](-0.5,6) -- (4.5,6);
\draw [dashed](-0.5,5) -- (6.5,5);
\draw [dashed](-0.5,4) -- (8.5,4);
\draw [dashed](-0.5,3) -- (8.5,3);
\draw [dashed](1.5,2) -- (8.5,2);
\draw [dashed](3.5,1) -- (8.5,1);
\draw [dashed](5.5,0) -- (8.5,0);
\draw [dashed](7.5,-1) -- (8.5,-1);
\draw [dashed](1,7.75) -- (1,2);
\draw [dashed](2,7.5) -- (2,1.5);
\draw [dashed](3,7) -- (3,1);
\draw [dashed](4,6.5) -- (4,0.5);
\draw [dashed](5,6) -- (5,0);
\draw [dashed](6,5.5) -- (6,-0.5);
\draw [dashed](7,5) -- (7,-1);
\draw [black] (2,7.05) node {$\times$};
\draw [black] (4,6.1) node {$\times$};
\draw [black] (6,5.13) node {$\times$};
\draw [black] (8.1,4.1) node {$\times$};
\draw [black] (8.1,-1.15) node {$\times$};
\draw [black] (2,1.9) node {$\odot$};
\draw [black] (4,0.95) node {$\odot$};
\draw [black] (6,-0.1) node {$\odot$};
\draw [black] (-0.1,8.15) node {$\odot$};
\draw [black] (-0.1,2.95) node {$\odot$};
\draw [black] (-0.1,4) node {$\odot$};
\draw [black] (-0.1,7) node {$\odot$};
\draw [black] (-0.1,5) node {$\odot$};
\draw [black] (-0.1,6) node {$\odot$};
\draw [black] (8.1,1) node {$\times$};
\draw [black] (8.1,3) node {$\times$};
\draw [black] (8.1,2) node {$\times$};
\draw [black] (8.1,0) node {$\times$};
\draw [black] (4.5,6.5) node {$W_1^\times$};
\draw [black] (-1,5) node {$W_2^\circ$};
\draw [black] (9,2) node {$W_2^\times$};
\draw [black] (4.5,0) node {$W_1^\circ$};
\draw [black] (0,8.5) node {{\small $(k,m)$}};
\draw [black] (8.5,-1.5) node {{\small $(k+2p,m-2p-1)$}};
\draw [black] (-0.3,2.1) node {{\small $(k,m-p-1)$}};
\draw [black] (8.5,4.7) node {{\small $(k+2p,m-p)$}};

\draw [black] (0,0.25) node {{\small $m$}};
\draw [black] (1.25,-1) node {{\small $k$}};
\draw [->,black](0,-1) -- (0,0);
\draw [->,black](0,-1) -- (1,-1);

\end{tikzpicture}
\end{center}
\caption{\emph{An  initial value problem for (\ref{eq:Discr2rA}). The black  staircase gives the initial values set for the even variables while the gray parallelogram gives the initial values for the odd variables (denoted by:  $\circ=\psi$ and $\times = \xi$).}} \label{fig:3}
\end{figure}
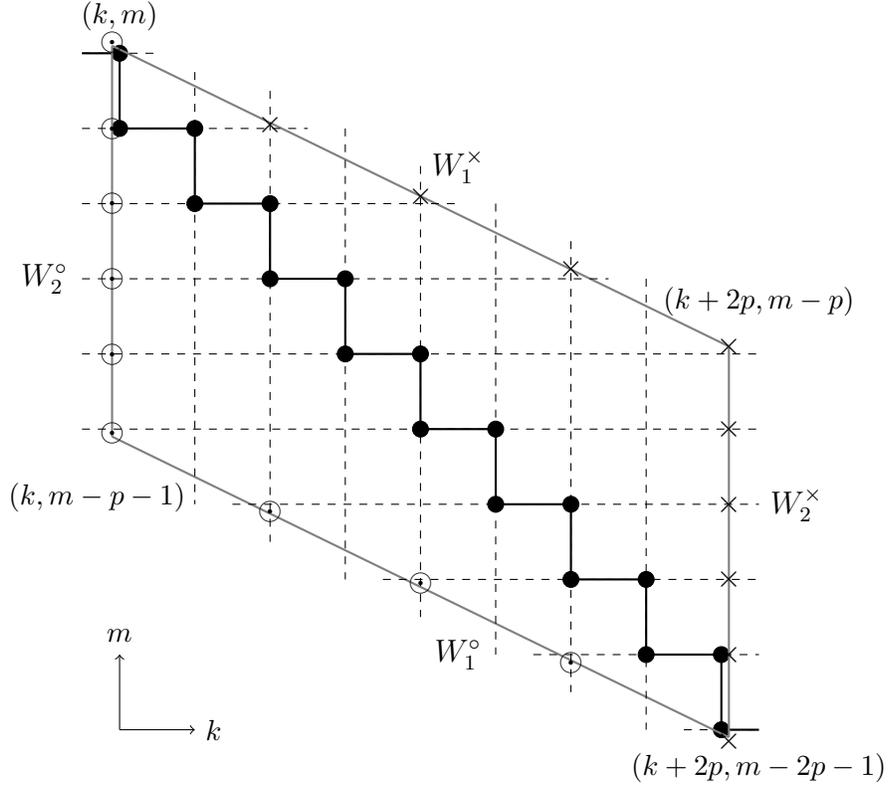

For the commutative Toda lattice (\ref{eq:Discr2r-TC}) one can solve an initial value problem with initial data given on the staircase $W_0$:
\begin{eqnarray}\label{eq:W0}
W_0=\{(k+n,m-n), (k+n,m-n-1) \, | \, n\in \{0,...,2p\}\}.
\end{eqnarray}
In the case of equations (\ref{eq:Discr2rA}) one also needs to define boundary odd variables. Taking some  $p\in {\Bbb N}$ we define a parallelogram $W$ (cf. Fig. {\ref{fig:3}}) with boundaries
\begin{eqnarray}\label{eq:W1p}
W_1^\times&=&\{(k+2n,m-n) \, | \, n\in \{1,...,p\}\}\\
\label{eq:W1m}
W_1^\circ&=&\{ (k+2n,m-n-p-1) \, | \, n\in \{0,...,p-1\}\}\\
\label{eq:W2}
W_2^\circ&=&\{(k,m-n)  \, | \, n\in \{0,...,p+1\}\}, \\
\label{eq:W3}
W_2^\times&=&\{(k+2p,m-n-p)  \, | \, n\in \{0,...,p+1\}\}.
\end{eqnarray}
The set of boundary variables $\psi_{km}^{(0)}$ are defined on $W^\circ=W_1^\circ \cup W_2^\circ$ and the boundary variables $\xi_{km}^{(0)}$  are defined on $W^\times=W_1^\times \cup W_2^\times$. It is easy to show, that any variable inside the parallelogram $W$ bounded by $W^\circ \cup W^\times$  can be uniquely expressed as a rational function of the even variables given on the staircase $W_0$ inscribed into the parallelogram $W$ and multi-linear functions of the odd variables on $W^\circ$ and $W^\times$.

Indeed, the system (\ref{eq:Discr2rA}) with such initial boundary conditions  can be solved by a finite sequence of iterations. For the first iteration we set all odd variables inside the parallelogram to zero and find the first approximation of even variables for all points  of $W$. Then, using the boundary conditions for odd variables, one can solve equations  (\ref{eq:elem_ev-xi}) to update the values of odd variables inside $W$. Starting from these data, we repeat the sequence of iterations. This sequence will stabilise after a finite number of steps since the solution ($q_{k_1,m_1}$, $\psi_{k_1,m_1}$, $\xi_{k_1,m_1}$), $(k_1,m_1)\in W$ is a multi-linear function of the odd boundary data. Thus one can define again an {\it Elimination map} \cite{MWX2} which will allow us to express the solutions as rational functions of the even initial data on the staircase $W_0$ inscribed into the parallelogram $W$ and multi-linear functions of the odd boundary data.

The mixed initial/boundary value problem presented here allows one to recover the standard staircase $W_0$ initial value problem \cite{Adl,SSP} by setting $\psi_{n,m}=\xi_{n,m}=0$.

\section{Conclusions}

In the present paper we studied integrable difference equations associated with  Grassmann extensions of the nonlinear Schr\"odinger equation. We constructed two elementary Darboux transformations. As a result, new Grassmann generalisations of the Toda lattice and the NLS dressing chain are obtained. We should mention here that the noncommutative NLS dressing chain we obtained is a nonlocal one - although the commutative reduction will lead to the standard NLS dressing chain \cite{SSP}. The Toda lattice dressing chain obtained here  differs from the known supersymmetric Toda chains \cite{Olshan,Evans,Bonora2,Ikeda}.

Furthermore, we obtained difference integrable systems as a  compatibility (Bianchi commutativity) of these Darboux transformations. Such systems can be viewed as Grassmann generalisations of the difference Toda and NLS equations. These lattice systems  have Lax pairs provided by the set of two consistent Darboux transformations. The corresponding B\"acklund transformations represent symmetries of the discrete (difference) systems.

For the two discrete systems obtained, we formulated the initial value problems: in the first case (when two Darboux transformations with NLS-type  dressing chains are combined) the well-posed {\it initial value problem} is defined (for both even and odd variables) by giving  an initial profile
along an infinite staircase within the $(k,m)$-lattice (Fig. \ref{fig:2}). In the second case, the well-posed {\it initial-boundary value problem} is defined for a parallelogram (of any size). The initial even variables should be given on a staircase inscribed in the parallelogram while the odd variables should be given on the boundary of the parallelogram (Fig. \ref{fig:3}). This differs from the commutative case where the initial value problem is globally well-posed on an infinite staircase. The bosonic limit  reconstructs the results of \cite{SSP}. However, one can still formulate a well-posed initial value problem on a stretched staircase \cite{Peter}, but in the bosonic limit it will not recover the standard staircase initial value problem for the Toda lattice \cite{Adl}.

The results obtained here can be developed in several directions: 1) to study the corresponding Yang-Baxter maps \cite{SS,PapagTong};  2) To derive the recursion operators and study the associates multi-Hamiltonian structures; 3) to study Darboux transformations and the corresponding difference equations with non-evolutionary dressing chains; 4) to study elementary Darboux transformations for Lax operators with nontrivial reduction groups; 5) this can be generalised to other integrable hierarchies.

\section*{Acknowledgements}

The authors have the pleasure to thank Prof. Qing Ping Liu, Dr. Pavlos Xenitidis, Dr Peter van der Kamp and Mr. Sotiris Konstantinou-Rizos  for numerous stimulating discussions. This is a part of a project supported by the  Leverhulme Trust. The work of A.V.M. is partially supported by the EPSRC (Grant EP/I038675/1 is acknowledged).

\label{last}

{\small

\begin{thebibliography}{99}\itemsep=-.2pc


\bibitem{Adler1} V. E. Adler, \emph{ Nonlinear chains and Painlev\'e equations}, Physica D {\bf 73} (1994) 335--351.

\bibitem{Adler2} V. E. Adler. \emph{Nonlinear superposition formula for Jordan NLS equations}, Phys. Lett. A {\bf 190} (1994) 53--58.

\bibitem{Adl} V. E. Adler. \emph{Classification of discrete integrable equations}, DSc thesis, Landau Institute, Chernogolovka (2010).

\bibitem{AntFor} M. Antonowicz, A. P. Fordy, \emph{Super-extensions of energy-dependent Schr\"odinger operators}, Comm. Math. Phys. {\bf 124} (1989), 487--500.

\bibitem{Ber} F. A. Berezin, {\it Introduction to superanalysis}, D. Reidel Publishing, Dordrecht/Boston/Lancaster/Tokyo (1987).


\bibitem{Bonora2} L. Bonora, S. Krivonos, A. Sorin, \emph{Towards the construction of $N=2$ supersymmetric integrable hierarchies}, Nucl. Phys. B 477 (1996), 835--854.

\bibitem{Chaish} M. Chaichian and P. P. Kulish, \emph{On the method of inverse scattering problem and B\"acklund transformations for supersymmetric equations}, Phys. Lett. B {\bf 78} (1978), 413--416.

\bibitem{Chen}
H. H. Chen, C. S. Liu,  \emph{B\"acklund transformation solutions of the Toda lattice
equation}, J. Math. Phys. {\bf 16} (1975) 1428--1430.



\bibitem{Ciesh2} J. L. Cie\'sli\'nski, \emph{ Algebraic construction of the Darboux matrix revisited}, J. Phys. A {\bf 42} (2009), no. 40, 404003.

\bibitem{DimMH}Dimakis, A., M\"uller-Hoissen, F., \emph{An algebraic scheme associated with the non-commutative
KP hierarchy and some of its extensions}, J. Phys. A {\bf 38} (2005), 5453--5505.

\bibitem{DimMH2}Dimakis, A., M\"uller-Hoissen, F., \emph{Solutions of matrix NLS systems and their discretizations: A unified treatment}, Inverse Problems {\bf 26} (2010) 095007.

\bibitem{DimMH3}Dimakis, A., M\"uller-Hoissen, F., \emph{Binary Darboux Transformations in Bidifferential Calculus and Integrable Reductions of Vacuum Einstein Equations}, SIGMA {\bf 9} (2013), 009, 31 pages.

\bibitem{DrSok*85}
Drinfel'd V. and Sokolov V. V., \emph{Lie Algebras and Equations
of Korteweg-de Vries Type}, Sov. J. Math. \textbf{30} (1985) 1975--2036.

\bibitem{Evans} J. Evans, T. Hollowood, \emph{Supersymmetric Toda field theories}, Nucl. Phys. B 352 (1991), 723--768.


\bibitem{Gelf2} I. M. Gelfand, S. I. Gelfand, V. S. Retakh, R. L. Wilson \emph{Quasideterminants}, Adv. Math. {\bf 25} (2005),56--141.


\bibitem{GrisPen} Grisaru, M. T., Penati, S., \emph{An integrable noncommutative version of the sine-Gordon
system}, Nucl. Phys. B {\bf 655} (2003), 250--276.



\bibitem{Gurs} M. G\"urses, \"O. O\u{g}uz, \emph{A super soliton connection}, Lett. Math. Phys. {\bf 11} (1986) 235--246.

\bibitem{HaTo} Hamanaka, M., Toda, K., {\it Towards noncommutative integrable systems}, Phys. Lett. A {\bf 316} (2003),
77--83.


\bibitem{Ikeda} K. Ikeda, \emph{A supersymmetric extension of the Toda lattice hierarchy}, Lett. Math. Phys. {\bf 14} (1987),
321--328.

\bibitem{IK} T. Inami and H. Kanno, \emph{Lie superalgebraic approach to super Toda lattice and generalized
super KdV equations}, Commun. Math. Phys. {\bf 136} (1991), 519--542.

\bibitem{Peter} Peter H. van der Kamp, \emph{Private Communication}.

\bibitem{SS} S. Konstantinou-Rizos and A. V. Mikhailov, \emph{Yang-Baxter maps and finite reduction groups with degenerated orbits}, E-print: {\tt arXiv:1205.4910}.

\bibitem{SSP} S. Konstantinou-Rizos, A. V. Mikhailov and P. Xenitidis, \emph{Reduction group and Darboux Transformations}, In preparation.

\bibitem{Kulish1} Kulish P. P., \emph{Quantum $osp(1|2)$-invariant nonlinear Schr\"odinger equation}, ICTP Preprint IC/85/39, Trieste (1985).

\bibitem{Kupersh} B. A. Kupershmidt, \emph{A super Korteweg-de-Vries equation} Phys. Lett. A {\bf 102} (1984), 213--215.


\bibitem{LLPPT} Lechtenfeld, O., Mazzanti, L., Penati, S., Popov, A. D., Tamassia, L., \emph{Integrable
noncommutative sine-Gordon model}, Nucl. Phys. B {\bf 705} (2005), 477--503.

\bibitem{Leites} D. A. Leites (ed), {\it Seminar on Supersymmetry}, Independent University Press, Moscow (2011) (in Russian).

\bibitem{Levi} D. Levi, \emph{Nonlinear differential difference equations as B\"acklund
transformations} J. Phys. A {\bf 14} (1981) 1083--1098.

\bibitem{LiNim} C. X. Li and J. J. C. Nimmo, \emph{Darboux transformations for a twisted derivation and quasideterminant solutions to the super KdV equation}, Proc. Royal Soc. A {\bf 466} (2010), 2471--2493.

\bibitem{LiuMan1} Q. P. Liu and M. Ma\~nas, \emph{Darboux transformations for the Manin-Radul supersymmetric KdV equation}, Phys. Lett. B {\bf 394} (1997), 337--342.

\bibitem{LiuMan2} Q. P. Liu and M. Ma\~nas, \emph{Crum transformation and Wronskian type solutions for supersymmetric KdV equation}, Phys. Lett. B {\bf 396} (1997), 133--140.

\bibitem{Manin} Yu. I. Manin and A. O. Radul, \emph{A supersymmetric extension of the Kadomtsev-Petviashvili hierarchy}, Commun. Math. Phys. {\bf 98} (1985), 65--77.

\bibitem{Mathieu} P. Mathieu, \emph{Supersymmetric extension of the Korteweg-de Vries equation}, J. Math. Phys. {\bf 29} (1988), 2499.

\bibitem{Matv} V. B. Matveev, M. A. Salle, \emph{Darboux transformations and solitons}, Springer Series in Nonlinear Dynamics {\bf 4}, Springer-Verlag, Berlin/Heidelberg/New-York (1991)

\bibitem{AVM} A. V. Mikhailov,   \emph{Integrability of Supersymmetric Generalization of the Classical Chiral Model in Two-dimensional Space-time}, JETPh Letters. {\bf 28} (1978) 554--558.



\bibitem{MWX2} A. V. Mikhailov, J. P. Wang and P. Xenitidis,  \emph{Cosymmetries and Nijenhuis recursion operators for difference equations}, Nonlinearity {\bf 24} (2011) 2079--2097.



\bibitem{MorPi} C. Morosi, L. Pizzocchero, \emph{A fully supersymmetric AKNS hierarchy}, Commun. Math. Phys. {\bf 176} (1996), 353--381.




\bibitem{Oevel} W. Oevel, Z. Popowicz, \emph{The bi-Hamiltonain structure of fully supersymmetric Korteweg-de-Vries systems},
Commun. Math. Phys. {\bf 139} (1991), 441-460.

\bibitem{OlvSok1}P. J. Olver and  V. V. Sokolov, \emph{Non-abelian integrable systems of the derivative nonlinear Schr\"odinger type}, Inverse Problems {\bf 14} (1998), no. 6, L5--L8.

\bibitem{OlvSok2}P. J. Olver and  V. V. Sokolov, \emph{Integrable evolution equations on associative algebras}, Comm. Math. Phys. {\bf 193} (1998), no. 2, 245--268.

\bibitem{Olshan} M. A. Olshanetsky, \emph{Supersymmetric two-dimensional Toda lattice}, Commun. Math. Phys. {\bf 88} (1983), 63--76.

\bibitem{PapagTong} V. G. Papageorgiou, A. G. Tongas, \emph{Yang-Baxter maps and multi–field
integrable lattice equations}, J. Phys. A {\bf 40} (2007) 12677--12690.

\bibitem{Pick} A. Pickering and Z. N. Zhu, \emph{New integrable lattice hierarchies}, Phys. Lett. A {\bf 349} (2006), 439--445.

\bibitem{Ziemek1} Z. Popowicz, \emph{The Extended Supersymmetrization of the Nonlinear Schr\"odinger Equation},
Phys. Lett. A {\bf 194} (1994), 375--379.







\bibitem{shabat} A. B. Shabat, \emph{ Inverse scattering problem for a system of differential equations}, Funct. Anal. Appl. {\bf 9} (1975), 244--247.

\bibitem{Toda} M. Toda, \emph{Theory of Nonlinear Lattices} (Springer Series of Solid-State Sciences {\bf 20}), Springer, Berlin-Heidelberg, 1981.

\bibitem{brown-bible}
Zakharov V. E., Manakov S., Novikov S.  and Pitaevskii L.,
\emph{Theory of Solitons: The Inverse Scattering Method},
Plenum, New York, 1984.

\bibitem{zm1}
Zakharov  V. E. and Mikhailov A. V., \emph{On The Integrability of Classical
Spinor Models in Two-dimensional Space-Time}, Commun. Math. Phys.
{\bf 74} (1980) 21--40.






\end{thebibliography}

}
\end{document}